# Strain-Tunable Magnetic Compensation Temperature of Epitaxial $Tb_3Fe_5O_{12}$ Thin Films

Yufei Li[1,*], Xihui Yang[2,*], Hua Bai[3], Mingzhi Wang[1], Dashuai Cheng[1], Cheng Song[3], Zhe Yuan[2], Yi Liu[2,†], Zhong Shi[1,‡]

[1]*Shanghai Key Laboratory of Special Artificial Microstructure Materials and Technology and Pohl Institute of Solid State Physics and School of Physics Science and Engineering, Tongji University, Shanghai 200092, China*

[2]*Center for Advanced Quantum Studies and Department of Physics, Beijing Normal University, Beijing 100875, China*

[3]*Key Laboratory of Advanced Materials, School of Materials Science and Engineering, Beijing Innovation Center for Future Chips, Tsinghua University, Beijing 100084, China*

**Abstract:**

High-quality rare-earth iron garnet (ReIG) $Tb_3Fe_5O_{12}$ (TbIG) thin films are epitaxially grown on a series of (111)-oriented garnet substrates with various lattice constants. The coherent growth induces a substrate-dependent in-plane tensile or compressive strain in the TbIG film. Measurements of the anomalous Hall-like effect (AHLE) in TbIG/Pt heterostructures show that the compensation temperature of TbIG films monotonically changes with the film strain. The strain results in a variation of the distances between magnetic atoms in the TbIG crystal and therefore the corresponding exchange interactions. The latter is explicitly calculated as a function of the lattice strain based on density functional theory, reproducing the observed experimental results. This work provides a versatile way to optimize ReIG-based spin-orbit torque devices.

**Introduction:**

Ferrimagnetic rare-earth iron garnets (ReIGs) containing coupled sublattices with inequivalent and antiparallel magnetic moments have attracted intensive research interest for their potential applications in the next generation of spintronic devices [1-4]. As typical ferrimagnets, the most common feature of ReIGs is the magnetic compensation point, where the magnetic moment of the rare earth ($M_{Re}$) is equal to that of iron ($M_{Fe}$) resulting in a vanishing remanent net magnetization. This characteristic of the ferrimagnets at the compensation point imparts upon them certain properties of antiferromagnetic materials, such as large coercive field [5], enhanced exchange coupling [6], and long spin coherence length [7,8]. Therefore, near the magnetic compensation point, ReIGs enable fast spin-orbit torque (SOT) switching [9], ultralow switching current [10] and offer promising properties such as low energy consumption, high efficiency, and robustness, making them compelling candidates for advanced spintronic applications. Meanwhile, the magnetization of sublattices can be individually distinguished using the magneto-optical effect [11-13] or electric transport effects such as spin Hall magnetoresistance [14,15]. This approach is promising for facilitating investigation of the exchange interactions within and among the sublattices, which primarily determine the compensation point, and provides insights into the underlying physics of ferrimagnetic materials [16-18].

To tune the compensation point, conventional ferrimagnetic alloys have been adjusted by altering the composition ratio [19,20]. In the context of ReIGs, the approaches to manipulate the compensation temperature entails replacing trivalent rare-earth ions (such as $Gd^{3+}$, $Tb^{3+}$, $Dy^{3+}$, and $Tm^{3+}$) at the relevant dodecahedron site, and/or replacing the iron ions at the tetrahedron/octahedron sites [21-23]. Very recently, a strain-tunable effect on the Néel temperature was observed in perovskite $YCrO_3$ thin films by inducing lattice strain from the substrate [24]. This demonstrates that strain has the ability to significantly alter the lattice constant, offering a powerful tool to control the compensation point without varying the chemical composition. Here, we implement this idea by performing a systematic study of the strain effect on the compensation temperature in ReIGs.

Terbium iron garnet ($Tb_3Fe_5O_{12}$, TbIG) has a larger positive magnetostriction constant ($\lambda_{111} = 1.2 \times 10^{-5}$) than the other ReIGs at room temperature [25,26]. Thus, a strain effect on the compensation temperature is expected to be tuned over a wide range in TbIG. In this paper, high-quality epitaxial TbIG thin films are grown on (111)-oriented single-crystalline garnet substrates, which introduce a systematic variation in the lattice strain of the TbIG films. By measuring the anomalous Hall-like effect (AHLE) in TbIG/Pt heterostructures as a function of temperature, we find that the compensation temperature ($T_{comp}$) of TbIG monotonically decreases with increasing lattice constant of the substrate. Using density functional theory, we explicitly calculate the exchange interactions between magnetic atoms in TbIG as a function of the lattice strain, which reproduces the observed variation in $T_{comp}$ in our experiment. For other rare earth iron garnets, where the rare earth ions have finite magnetic moments, the strain-tunable effect on the compensation temperature is also expected. The current results provide deeper insight for controlling the magnetic properties in ferrimagnetic materials and are helpful for facilitating the application of ReIG-based SOT devices.

**Sample preparation:**

Five kinds of (111)-oriented single-crystalline garnet substrates, including $Gd_3Ga_5O_{12}$ (GGG), $Y_3Sc_2Ga_3O_{12}$ (YSGG), $Gd_{2.6}Ca_{0.4}Ga_{4.1}Mg_{0.25}Zr_{0.65}O_{12}$ (SGGG), $Nd_3Ga_5O_{12}$ (NGG), and $Gd_3Sc_2Ga_3O_{12}$ (GSGG), were selected for deposition of 20-nm-thick TbIG films by pulsed laser deposition (PLD). Every batch of five substrates was precleaned, subsequently affixed to a circular sample stage at the same time and evenly distributed in a circle. The sample stage was then transferred into the PLD chamber of a PLD-sputtering linked system. The base pressure of the PLD chamber was less than $2\times10^{-6}$ Pa, and the substrates were preheated to and kept at 750 °C for 1 hour before film deposition. During deposition, a KrF ($\lambda$ = 248 nm) excimer pulsed laser with a laser fluence ~1.22 J/cm$^2$ was used to strike a stoichiometric polycrystalline TbIG target with a repetition rate of 4 Hz in a PLD chamber atmosphere of 3.0 Pa oxygen. To ensure the homogeneous growth of samples in one batch, the sample stage rotates clockwise at a certain rate $\omega_1$ during deposition, with the laser plasma plume swept over the center of each substrate (see more details in Figure S1(a) of Supplementary Material [27]). Twenty-nanometer-thick TbIG films were simultaneously deposited on the five substrates with a growth rate of 0.33 Å/s. The batch of films was then crystalized by in-situ annealing at 750 °C for another 2 hours in a pure oxygen atmosphere of 6×10$^4$ Pa. After that, the samples were cooled to room temperature and transferred to the magnetron sputtering chamber without breaking vacuum. A 3-nm-thick platinum (Pt) layer was deposited on each TbIG film at room temperature by magnetron sputtering with a 0.4-Pa Ar pressure. The deposition rate of Pt was 0.68 Å/s. UV lithography and ion beam etching were employed on these substrate/TbIG (20 nm)/Pt (3 nm) samples to fabricate identical Hall-bar patterns for the subsequent electric transport measurements. Another batch of 20-nm-thick TbIG films without Pt capping layers was also grown for crystallographic characterization and magnetometry measurements.

**Electric transport and magnetostatic measurements:**

We first carried out transport measurements of the TbIG (20 nm)/Pt (3

nm) heterostructures on the five different substrates. Fig. 1(a) depicts a schematic of the device geometry. The driven current $I$ flowed into the Pt layer in the $x$ direction, and the magnetic field $H$ was applied in the $z$ direction. Then, the Hall voltage was detected in the $y$ direction by a VersaLab system in the temperature range from 70 K to 300 K. In TbIG/Pt heterostructure, the magnetic proximity effect in Pt layer, induced by the nearby magnetic TbIG layer via exchange interaction, can generate an anomalous Hall voltage. In addition, the inverse spin Hall effect also leads to an AHLE signal depending on the spin mixing conductance of the TbIG/Pt interface [28,29]. All the AHLE signals are essentially determined by the magnetization of $Fe^{3+}$ in TbIG since the 3$d$ electrons are much more extended than the 4$f$ electrons in $Tb^{3+}$. Therefore, the observed anomalous Hall-like signal primarily reflects the magnetic properties of $Fe^{3+}$.

Taking NGG/TbIG/Pt as an example, the magnetic field dependence of the Hall resistance ($R_H$) at various temperatures is plotted in Fig. 1(b). The $R_H$ - $H$ loops of NGG/TbIG/Pt exhibit square shapes with sharp jumps at all temperatures except for 175 K due to the well-established PMA in TbIG grown on the NGG substrate. This is similar to previous works on ReIG/Pt heterostructures with PMA [11,28,30-34]. Therefore, we can extract the anomalous Hall-like spin Hall resistance ($R_{AH}$) by calculating $R_{AH} = (R_H^+ - R_H^-)/2$, where $R_H^+$ and $R_H^-$ are the two intersection points with the $y$-axis obtained by extrapolating the $R_H$ - $H$ loop at positive and negative saturated fields, as marked by the dashed lines in Fig. 1(b). As the temperature increases, $R_{AH}$ abruptly changes sign from positive to negative at 175 K. This is because as the temperature increases across $T_{comp}$, the magnetization of $Fe^{3+}$ becomes larger than that of $Tb^{3+}$ such that the magnetic moments on $Fe^{3+}$ reverses to guarantee the net magnetization to be aligned with the external magnetic field. Thus, the measured Hall voltage reverses at $T_{comp}$ of TbIG [35]. In particular, when the magnetization in the ferrimagnet is just compensated at $T_{comp}$, the system has zero net magnetization resulting in the divergent coercive field.

It is worth noting that the sign change of $R_{AH}$ occurs in a very narrow temperature window. As shown in Fig. 1(b), $R_{AH}$ approaches zero at 175 K, exhibiting a straight line in the $R_H$ - $H$ curve. However, a negative (positive) $R_{AH}$ with a square loop resurges at 176 K (174 K), only 1 K away from $T_{comp}$. Therefore, the sign change of $R_{AH}$ for the NGG/TbIG/Pt heterostructure can be used to determine the $T_{comp}$ of TbIG with a precision of 1 K, as shown by the red circles in Fig. 1(c). The coercive field of the TbIG film on NGG is plotted as the black triangles in Fig. 1(c), which tends to diverge at 175 K, in agreement with the $R_{AH}$ measurement.

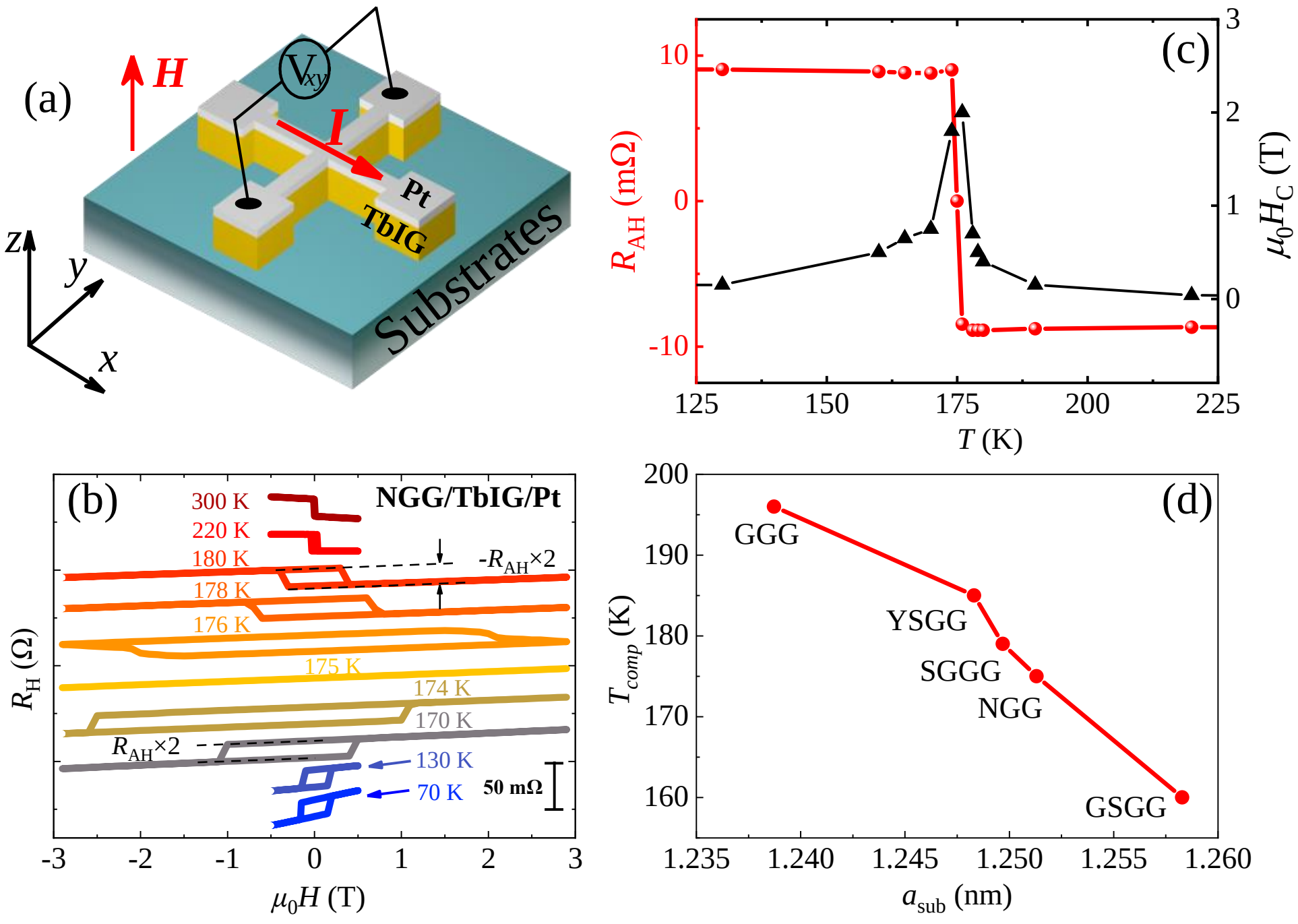

FIG. 1. (a) Schematic geometry of the transport measurement on the TbIG/Pt heterostructure. (b) Hall resistance ($R_H$) versus applied magnetic field for TbIG/Pt on the NGG substrate at various temperatures. $R_{AH}$ is defined as half of the $R_H$ jump at the $y$-axis, as indicated by the black dashed lines at 170 K and 180 K. (c) $R_{AH}$ (circles) and coercivity ($\mu_0 H_C$) (triangles) as a function of temperature for NGG/TbIG/Pt extracted from the transport measurement. (d) $T_{comp}$ for TbIG on different substrates as a function of the lattice constant of the garnet substrate.

To verify the $T_{comp}$ determined by the AHLE, we further measured the temperature-dependent magnetization and coercivity of the TbIG film on NGG by a vibrating sample magnetometer (VSM, see Figure S2 in Supplementary Material [27]). However, due to the near-zero net magnetization of the TbIG film itself in the vicinity of $T_{comp}$ and the substantial paramagnetic signal originating from the NGG substrate, it is a challenge to extract the net magnetization data from TbIG film within the temperature close to $T_{comp}$. Therefore, the sign change temperature of $R_{AH}$ is a better estimate for the $T_{comp}$ of TbIG due to its sensitivity and accuracy.

On different garnet substrates (see more details in Figure S3 of Supplementary Material [27]), $T_{comp}$ is 160 K (GSGG), 175 K (NGG), 179 K (SGGG), 185 K (YSGG), and 196 K (GGG), as shown in Fig. 1(d), where the $T_{comp}$ of TbIG is found to monotonically decrease with increasing lattice constant of the substrate.

**Structure and strain characterization:**

$T_{comp}$ is primarily dominated by the exchange interactions within and among the magnetic Tb and Fe ions in TbIG, which is varied by the lattice distortion due to the different substrates. To determine the relationship between the $T_{comp}$ of TbIG and the lattice constant of the relevant substrates, we investigated the morphology of bare TbIG films without Pt capping layers. The film thickness was characterized at room temperature by X-ray reflection (XRR), which confirmed the consistent thickness of 20 nm for all the films on the five different substrates, and the root-mean-square surface roughness was less than 2 Å as measured by atomic force microscopy. To examine the epitaxial growth and crystalline ordering, we performed cross-sectional high-angle annular dark-field (HAADF) imaging of the TbIG film on GGG (111) viewed along the $[1\bar{1}0]$ direction using scanning transmission electron microscopy (STEM), as shown in Fig. 2(a). The STEM image shows high-quality epitaxy of the TbIG film throughout the whole region. The yellow dashed line indicates the GGG/TbIG atomically sharp interface with perfect continuation of the garnet lattice, without any detectable dislocations or other defects. The sharp interface and homogeneity are further confirmed from the element distribution analysis by energy dispersive X-ray spectroscopy (see more details in Figure S1(b) to (d) and Table S1 of Supplementary Material [27]).

A conventional parameter $\eta$ is usually adopted to characterize the lattice strain and deformation induced by the lattice mismatch between the TbIG film and substrate as $\eta = \frac{a_{\mathrm{sub}} - a_{\mathrm{TbIG}}}{a_{\mathrm{TbIG}}}$, where $a_{\mathrm{sub}}$ and $a_{\mathrm{TbIG}}$ are the lattice constants of the substrate and TbIG bulk material, respectively [26,36-38] (here, $a_{\mathrm{TbIG}}$ = 1.2460 nm from Ref. [26]). To describe the distortion of TbIG, we must also consider the out-of-plane strain in the TbIG film in addition to the in-plane strain parameter $\eta$. As shown in Fig. 2(b), the XRD curves taken from the TbIG films on various substrates show two main (444) Bragg peaks for the TbIG films (marked by up arrows) and garnet substrates. Laue oscillations around film peaks indicate the precise spatial periodicity of the epitaxial TbIG lattice, and smooth interfaces on the corresponding substrates. The Bragg peaks of the substrates shift towards the right-hand side from GSGG to GGG, corresponding to the monotonically decreasing lattice spacing $d_{444}^{\mathrm{sub}}$. The lattice constants $a_{\mathrm{sub}}$ are derived and listed in Table I. In contrast, the (444) Bragg peaks of the TbIG films (up arrows) shift towards the left-hand side to a lower angle with decreasing substrate lattice spacing from GSGG to GGG. In particular, the peak of TbIG on GSGG shows a distinct crossover as it moves to the right-hand side of the GSGG peak. The diffraction peaks suggest that for the TbIG films, there are opposite variations in the in-plane lattice spacing controlled by the substrate and the out-of-plane lattice spacing, as listed by $d_{444}^{\mathrm{TbIG}}$ in Table I. The epitaxial growth of TbIG exhibits anisotropic distortion compared with the bulk structure. Similar phenomena have also been observed in previous reports [38,39]. Thus, the parameter $\eta$ is not sufficient to characterize the lattice distortion or distinguish the in-plane and out-of-plane strain of TbIG.

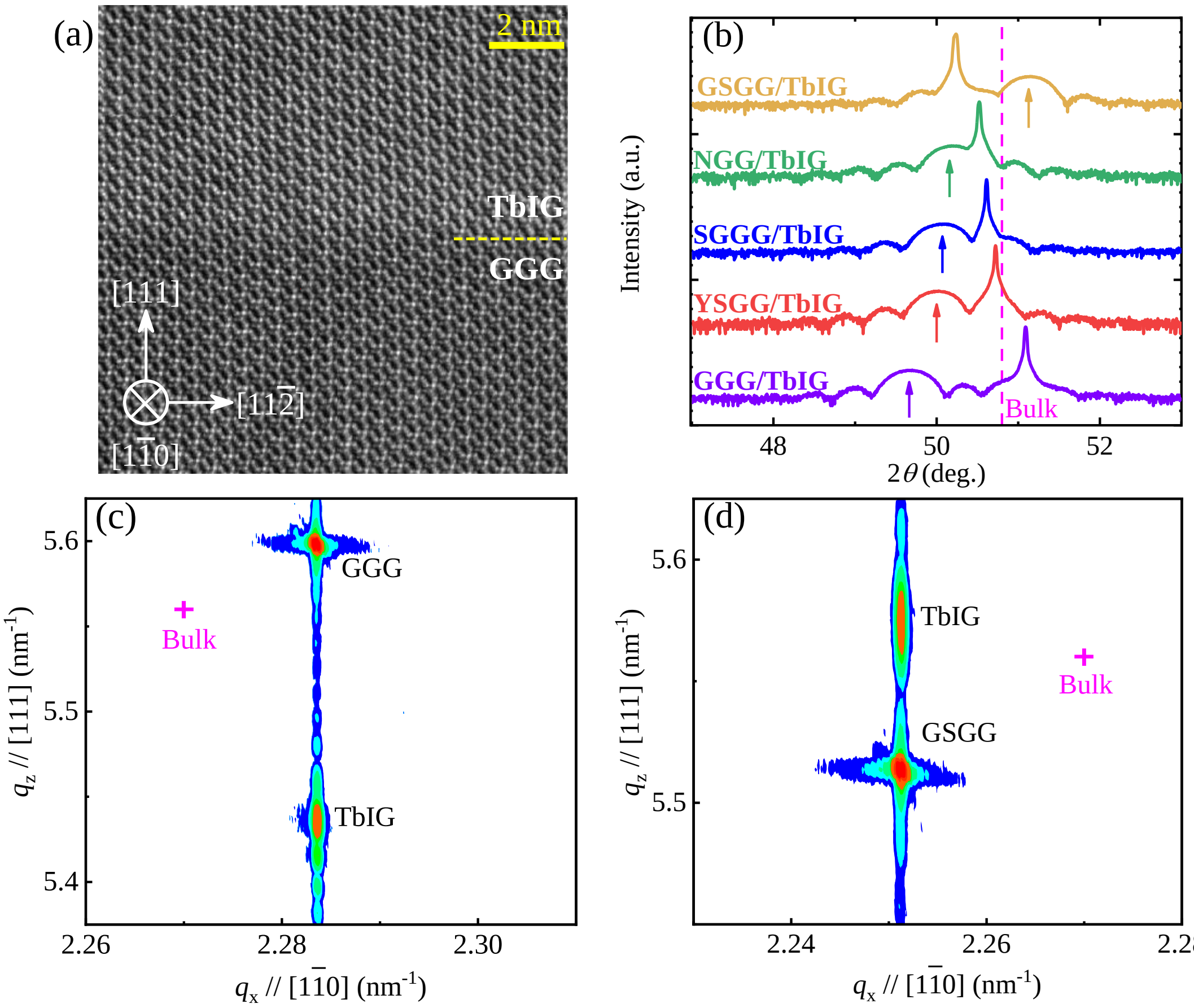

FIG. 2. (a) Cross-sectional HAADF-STEM image of a TbIG film grown on (111)-oriented GGG, viewed along the [1$\bar{1}$0] direction. (b) X-ray diffraction patterns of 20-nm-thick TbIG films deposited on (111)-oriented GGG, YSGG, SGGG, NGG, and GSGG. Curves are vertically offset for clarification. The up arrows represent the position of the (444) Bragg peaks of TbIG films. The dashed line indicates the (444) Bragg peak of bulk TbIG [26]. Reciprocal space mappings of the asymmetric scan around the (624) peaks of TbIG films on GGG (c) and GSGG (d) substrates. Crosses show the calculated peak positions for bulk cubic TbIG.

To elucidate the lattice distortion and strain distribution, reciprocal space mapping (RSM) was performed around the (624) asymmetric peaks of TbIG on GGG and GSGG, corresponding to the lower and upper bounds of the in-plane lattice constant in the series of garnet substrates, and the results are shown in Figs. 2(c) and (d), respectively. This RSM plot provides rich structural information of the TbIG films on the substrates. First, the lattice spacings $d_{220}$ and $d_{444}$ calculated from the RSM plot of GGG (GSGG) are in good agreement with the lattice constant of $a_{GGG}$ = 1.238 nm ($a_{GSGG}$ = 1.257 nm) for the cubic GGG (GSGG) crystal. Second, the RSMs reveal that the $q_x$ // $[1\bar{1}0]$ orientation of the TbIG film is nearly identical to those of the GGG and GSGG substrates, suggesting that the TbIG films grow coherently within the plane and adopt the lattice constant of the substrate. Since GGG and GSGG are the lower and upper limits for the lattice constant of the five substrates, it is reasonable to believe that all the TbIG films are in-plane fully strained to YSGG, SGGG and NGG [38,39]. Moreover, the RSM plot of the TbIG layer on GGG (GSGG) is below (above) that of the substrate along the $q_z$ // [111] direction, indicating that the out-of-plane lattice spacing is larger (smaller) than that of the substrate. Therefore, the TbIG film is deformed with a compressive (tensile) strain on the GGG (GSGG) substrate, leading to an increase (decrease) in the out-of-plane lattice spacing ($d_{444}$) for the film.

To explicate the variation of out-of-plane lattice distance, we prepared TbIG films with different thickness on GSGG substrate, of which the (444) diffraction peaks are shown in Fig. S4 of Supplementary Material [27]. With the film thickness increasing, the out-of-plane lattice spacing $d_{444}$ is gradually approaching the bulk value. In our transport measurement, the AHLE signals are essentially determined by the magnetization of $Fe^{3+}$ in the TbIG layer adjacent to the Pt, which would have the least out-of-plane distortion. Therefore, without loss of generality, the structure of the topmost TbIG layer is theoretically modelled in the following as a fully strained in-plane lattice that matches the corresponding substrate, while the out-of-plane lattice spacing is assumed to be that of bulk TbIG.

TABLE I. Structural parameters and compensation temperatures for the TbIG thin films grown on different substrates. The lattice constant of the substrates ($a_{\text{sub}}$) and the out-of-plane lattice spacing of TbIG ($d_{444}^{\text{TbIG}}$) are determined by XRD. The parameter $\eta$ is derived from $a_{\text{sub}}$ and the bulk TbIG lattice constant ($a$ = 1.2460 nm) [26]. $T_{comp}$ is determined from the $R_{AH}$ - $T$ curves.

| Substrate | $a_{\text{sub}}$ (nm) | $\eta$ (%) | $d_{444}^{\text{TbIG}}$ (nm) | $T_{comp}$ (K) |
|---|---|---|---|---|
| GGG | 1.23757 | -0.67658 | 0.18338 | 196 |
| YSGG | 1.24599 | 0.00872 | 0.1822 | 185 |
| SGGG | 1.24852 | 0.20249 | 0.18201 | 179 |
| NGG | 1.25049 | 0.35997 | 0.18165 | 175 |
| GSGG | 1.25711 | 0.89202 | 0.17843 | 160 |

**Theoretical Calculations**: Typically the compensation temperature of a ferrimagnet is determined by the competition between the magnetization of sublattices in mean-field theory [21,40,41], where the most important parameters are the exchange interactions within and among each sublattice. To investigate the effects of the lattice structure on $T_{comp}$, we first calculate the exchange interactions in ferrimagnetic TbIG based on density functional theory. In a crystal cell of TbIG, there are three types of magnetic ions associated with different coordination polyhedrons of oxygen ions: 24 Tb ions in dodecahedrons (the *c* sites), 16 $Fe^{O}$ ions in octahedrons (the *a* sites) and 24 $Fe^{T}$ ions in tetrahedrons (the *d* sites). Thus, there are five exchange parameters among the sublattices [42,43]: the $J_{ad}$ between the two nearest Fe ions that belong to the two types of Fe with antiparallel atomic magnetizations, the $J_{aa}$ and $J_{dd}$ between the nearest $Fe^{O}$ at the *a* sites and the nearest $Fe^{T}$ at the *d* sites, respectively, and the $J_{ac}$ and $J_{dc}$ between the rare earth atom and the two types of Fe ions, respectively. We can map the total energy as

$$E_{\text{tot}} = E_0 + \frac{1}{2}\sum_{i\neq j} J_{ij}\boldsymbol{\mu}_i \cdot \boldsymbol{\mu}_j, \tag{1}$$

where $E_0$ is the magnetization-independent energy of the system and $\boldsymbol{\mu}_i$ is the atomic magnetization of the *i*-th magnetic ion. The exchange interactions $J_{ij}$ are considered short-ranged, and only the five types discussed above are taken into account. We artificially construct 8 inequivalent collinear spin configurations with atomic magnetizations flipped at different sites (see more details in Figure S5 in Supplementary Material [27]). The total energies of these configurations are calculated from first principles (see more details in Table S2 in Supplementary Material [27]), and a least-squares fit [42-45] using Eq. (1) determines the values of the exchange interactions.

The total energy calculation is carried out using the Vienna *ab initio* simulation package (VASP) [46,47]. The Perdew-Burke-Ernzerhof (PBE) [48] functional is employed to describe the exchange and correlation effect. A PBE version of the all-electronic projector augmented wave (PAW) method [49] is adopted with the $3d^6 4s^2$ configuration of Fe, $4f^9 6s^2$ configuration of Tb and $2s^2 2p^4$ configuration of O treated as valence electrons. An on-site Coulomb correction is employed with the Hubbard $U$ and Hund's $J$ parameters chosen as $U$-$J$=4.7 eV for Fe ions and $U$-$J$=3.3 eV for Tb ions [50]. A gamma-centered $k$-point mesh of $11\times11\times11$ is used to sample the Brillouin zone, and the cut-off energy of the plane wave basis vector is chosen as 700 eV. The above parameters have been tested for convergence. The choice of the Hubbard $U$ and Hund's $J$ parameters does not influence our magnetic ground state, and the calculated magnetization is found to be insensitive to the specific $U$-$J$ values.

The calculated exchange interactions as a function of the bond length are plotted using solid symbols in Fig. 3. Shaded areas mark corresponding value ranges according to computational uncertainties. It is assumed that TbIG maintains the bcc crystal structure, but its lattice constant is chosen to match that of the substrate. Our calculated exchange coefficients (black cross) agree reasonably well with the experimentally fitted values of bulk TbIG [51] where the most important $J_{ad}$ = 2.62 meV was extracted from experimental measurement compared to our calculated value of 2.59 meV. When the lattice constant of the substrates, as listed in Table I, is imposed on the TbIG lattice, TbIG becomes compressed (on the GGG substrate) and expanded (on the other four substrates) accordingly. Then, the calculated exchange interaction coefficients, as shown by the solid symbols in Fig. 3, monotonically decrease with the bond length or lattice constant. It is worth mentioning that all the relevant bond angles are not changed with the bcc structure maintained.

It is straightforward to divide these exchange coupling into two categories: $Fe^{3+}$-$Fe^{3+}$ (or $3d^5 - 3d^5$) coupling, namely $J_{aa}$, $J_{dd}$, and $J_{ad}$ as shown in Fig. 3 (a-c) (left column), and $Fe^{3+}$-$Tb^{3+}$ (or $3d^5 - 4f^8$) coupling, namely $J_{ac}$ and $J_{dc}$ as shown in Fig. 3 (d-e) (right). The former ones are antiferromagnetic (AFM) superexchange coupling according to the Goodenough-Kanamori-Anderson (GKA) rule [52-54], expressed in terms of positive exchange coefficients in our formalism. Their magnitude is determined by the corresponding bond lengths. Increasing the lattice constant leads to decrease in all three of them. For the latter category, our calculated results show that the exchange type is more intuitive: with the atomic magnetizations parallel with each other, the exchange, $J_{ac}$, between $Fe^{O}$ and Tb ions is ferromagnetic (FM), while the exchange, $J_{dc}$, between $Fe^T$ and Tb ions is AFM with their magnetizations anti-parallel. The FM $J_{ac}$ also decreases linearly with increasing bond length, but the AFM $J_{dc}$ keeps almost constant. This insensitivity to bond length may result from its significantly short bond length (the shortest one among the five), the overlap integral of electronic orbitals that determines the exchange strength is nearly unchanged as the lattice size varies in the region we studied.

Our calculated exchange coefficients are reasonably consistent with the reported values in literature. The dominant positive value of $J_{ad}$ results in the antiferromagnetic coupling of $Fe^O$ and $Fe^T$ ions, as in many iron garnets [42,50,51,55-57]. In addition, our calculated $J_{dc}$ is positive in agreement with the positive values in these experimental and theoretical studies for GdIG [42,50,51,55-57], NdIG [50] and TbIG [51]. For the exchange coefficient $J_{ac}$, we find a small negative value indicating the weak ferromagnetic coupling between $Fe^O$ and Tb ions. The negative sign is the same to the calculated $J_{ac}$ in GdIG, but is opposite to the extracted value for TbIG [51]. The difference can be attributed to the approximation applied in fitting the experimental susceptibility that $J_{ad}$, $J_{aa}$ and $J_{dd}$ were assumed to be the same as in yttrium iron garnet (YIG). In fact, both the studies of YIG [45] and GdIG [42] demonstrate that the exchange coefficients sensitively depend on the lattice constant. Therefore, quantitative values of the exchange coefficients in ReIG must be carefully treated by taking the corresponding lattice structure and electronic properties into account.

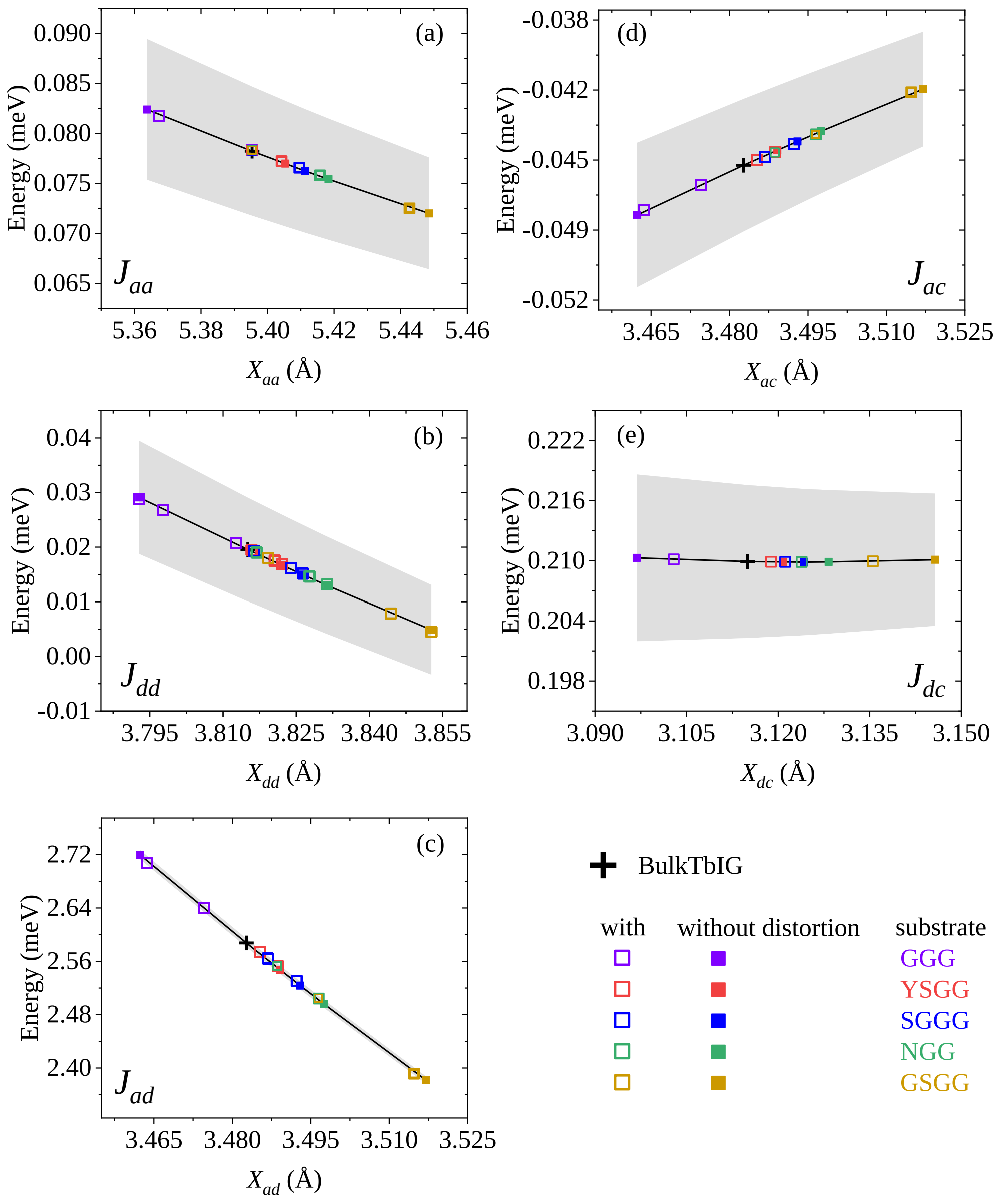

FIG. 3 Calculated exchange interactions between nearest atomic magnetizations in TbIG as a function of the corresponding bond length. Solid symbols present the values for bcc TbIG with the lattice constant the same as the substrates. Open symbols mark the values interpolated from the solid symbols corresponding to the changed bond lengths when lattice distortion is taken into account. The shaded areas mark the standard deviations of the calculated values.

We further take into account the anisotropic lattice distortion: the in-plane lattice constant is the same as that of the substrate, but the out-of-plane layer distance along the film normal direction remains the same as that in the bulk TbIG. For the five different substrates, we determine their changed bond lengths for every pair of magnetic ions. Interpolating along the lines in Fig. 3 using these bond lengths, we obtain modified exchange interactions (open symbols) due to anisotropic lattice distortion. For each type of exchange, there can be in principle as many different values as the number of pairs. In the simplest case, as for $J_{aa}$, every $Fe^{O}$ ion has 8 nearest neighbors of the same sublattice in the original bcc structure. Lattice distortion divides these neighboring pairs into two categories, as shown in Fig. 3(a): one category has similar bond length as that in the substrate due to the in-plane lattice matching, while the other has the same bond length as that in bulk TbIG resulting from the out-of-plane layer distance. Therefore, as shown in Fig. 3(a), there are five open symbols along the $J_{aa}$ value line, and the other five superposed on the cross stand for bulk TbIG. The situation is significantly different for other exchange interaction parameters, as there can be more bond lengths for other pairs of magnetic ions.

With the exchange interaction of every ion pair determined, we obtain the average exchange parameters for each sublattice and calculate the compensation temperature, $T_{comp}$, within mean-field theory [21,41]. Each atomic magnetization of sublattices $Fe^{O}$, $Fe^{T}$, and Tb is subject to the mean exchange field from the magnetizations of its neighbors:

$$H_a = -\left[\tilde{J}_{aa} \cdot M_a(T) + \frac{1}{2}\tilde{J}_{ad} \cdot M_d(T) + \frac{1}{2}\tilde{J}_{ac} \cdot M_c(T)\right], \quad (2)$$

$$H_d = -\left[\frac{1}{3}\widetilde{J}_{dd}\cdot M_d(T) + \frac{1}{2}\widetilde{J}_{ad}\cdot M_a(T) + \frac{1}{6}\widetilde{J}_{dc}\cdot M_c(T)\right], \quad (3)$$

$$H_c = -\left[\frac{1}{2}\widetilde{J}_{ac}\cdot M_a(T) + \frac{1}{6}\widetilde{J}_{dc}\cdot M_d(T)\right]. \quad (4)$$

Here, all the parameters $\widetilde{J}$ represent the average values of the different exchange interactions, and $H_\xi$ and $M_\xi(T)$ are the effective field and the total magnetization of the $Fe^O$ ($\xi = a$), $Fe^T$ ($\xi = d$) and Tb ($\xi = c$) sublattices, respectively. The magnetization as a function of temperature is expressed in terms of the Brillouin function $B_S(x)$:

$$M_\xi(T) = N_\xi|\mu_\xi|\cdot B_S(x_\xi), \quad (5)$$

$$B_S(x_\xi) = \frac{2S_\xi+1}{2S_\xi}\coth\left(\frac{2S_\xi+1}{2S_\xi}x_\xi\right) - \frac{1}{2S_\xi}\coth\left(\frac{1}{2S_\xi}x_\xi\right), \quad (6)$$

$$x_\xi = |\mu_\xi|\cdot H_\xi/(k_B\cdot T), \quad (7)$$

where $N_\xi$ is the number of ions within each sublattice in the unit cell ( $N_a = 8$, $N_d = N_c = 12$ ), $\mu_\xi$ is the atomic magnetization ($\mu_a = 5$, $\mu_d = -5$, $\mu_c = 6$, in the unit of the Bohr magneton $\mu_B$), $S_a = S_d = 5/2$ is the spin of Fe ions, and $S_c = 3$ for Tb ions. It is worth mentioning that the single-ion anisotropy field on Tb ions of approximately 1 T (estimated from the saturation field along the hard axis in Fig. S6) is neglected in our mean-field theory since it is much smaller than the exchange fields.

Solving Eqs. (2)-(7) self-consistently, we can calculate the magnetization $M_\xi(T)$ at every temperature. The total magnetization of the system,

$$M_{tot}(T) = M_a(T) + M_d(T) + M_c(T), \tag{8}$$

is thus determined. Choosing certain values of the exchange interactions (from the shaded areas in Fig. 3), we show in Fig. 4(a) the calculated $M_{tot}(T)$ for varying in-plane lattice constants imposed by the corresponding substrates. $T_{comp}$ is determined as the temperature at which $M_{tot}(T)$ vanishes. The results are plotted in Fig. 4(b). The shadow illustrates the uncertainties in evaluating the exchange interaction coefficients. By quantitatively analyzing the contribution from each exchange interaction to $T_{comp}$, we find that $J_{ad}$ only has marginal influence on the variation of $T_{comp}$ despite of its largest magnitude among the exchange parameters. In addition, the influence from $J_{aa}$ and $J_{dd}$ are also negligible. On the contrary, the dominant influence originates from $J_{ac}$ and $J_{dc}$, whose uncertainties basically determine the shadowed range in Fig. 4(b). This is because $T_{comp}$ arises from the competition of the magnetization of Fe and that of Tb, and the exchange interactions between Fe and Tb ($J_{ac}$ and $J_{dc}$) determines their relative magnitude of magnetization. The experimentally observed monotonic decrease in $T_{comp}$ with increasing lattice constant $a_{\mathrm{sub}}$, as plotted by the red curve, falls almost completely into the range predicted by mean-field model with an offset on the absolute values of $T_{comp}$.

For bulk TbIG, as marked by the pink dashed line, our mean-field simulation yields a range, 256 ~ 286 K, for $T_{comp}$. Compared to the experimentally measured $T_{comp}$ = 246 K [58], this is overestimation which can be understood as following. Due to the lack of magnon-magnon interaction, which lowers the total energy of the system and in turn softens the magnon spectrum with increasing temperature, the magnetization decreases faster than theoretically predicted by mean-field theory.

As the measured $T_{comp}$ on TbIG thin films is much lower than the bulk value [11,32], when comparing the theoretical $T_{comp}$ to the experimental values, we offset the right axis for the calculated values to plot them together with the measured values (left axis) in the same temperature range, i.e., a window of 50 K for both the theoretical results (245~295 K) and experimental values (150~200 K) in Fig. 4(b). The measured $T_{comp}$ values of the TbIG films grown on different substrates (red circles) fall into the theoretically calculated range (the shaded area). Satisfactory agreement is found between them, indicating that the main physical mechanism behind the monotonic decrease in the $T_{comp}$ of TbIG with increasing in-plane lattice constant of the substrate is the correspondingly varying exchange interaction. As the substrate changes from GGG to GSGG (lattice size changes from smaller to larger than that of bulk TbIG), the exchange interaction decreases as the bond lengths increase (except for $J_{dc}$, which is nearly unchanged). In particular, as the lattice constant increases, the exchange field exerted on Tb ions rapidly decreases, leading to a significant decrease in $T_{comp}$. This may help optimize the spin-orbit torque and facilitate the related applications of ferrimagnetic materials near the compensation temperature [20].

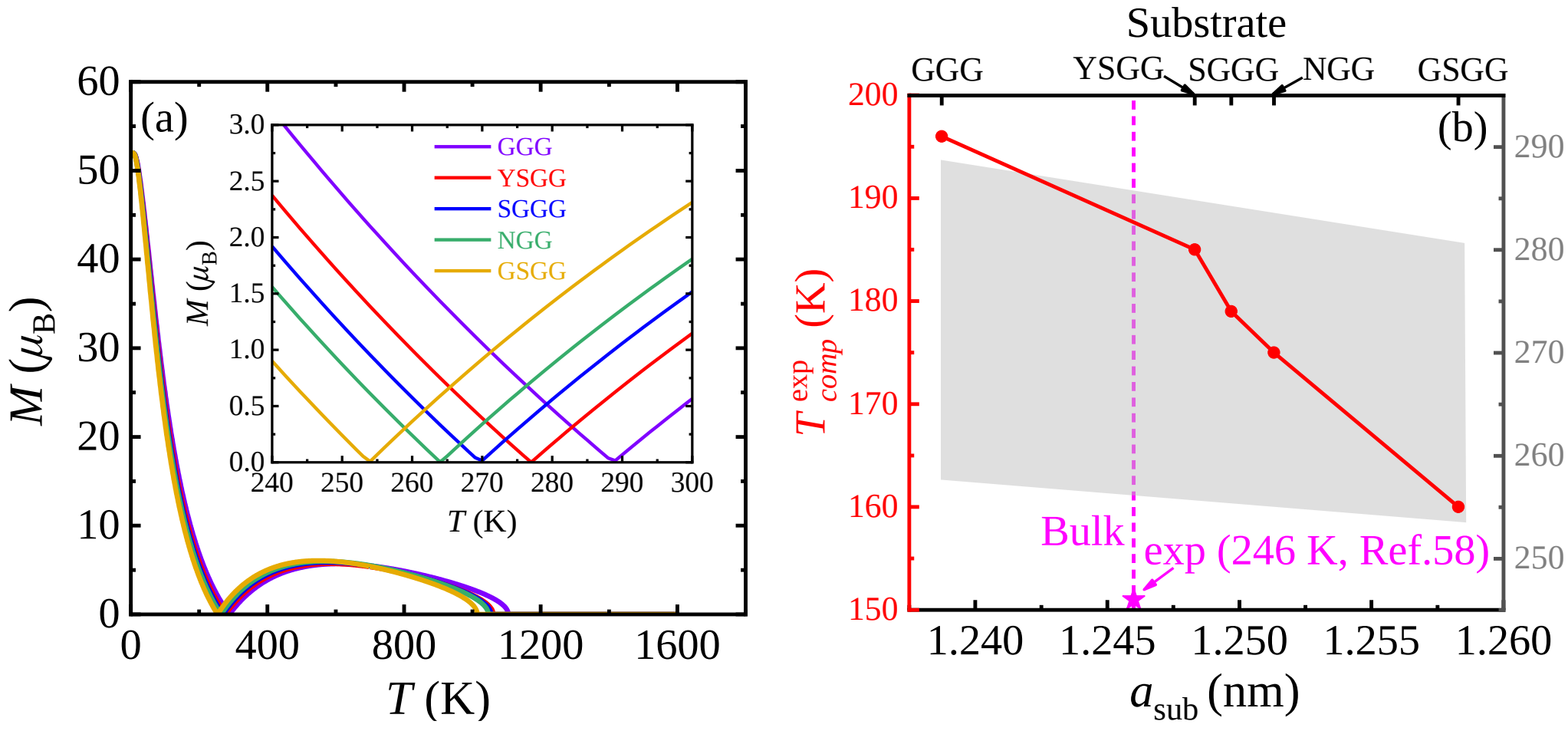

FIG. 4 (a) Calculated total magnetization as a function of temperature for distorted TbIG with different in-plane lattice constants as in the substrates. The inset enlarges the curves around the compensation temperature. (b) Measured (left axis) and calculated (right axis) $T_{comp}$ for TbIG as a function of the (in-plane) substrate lattice constant. The pentagram indicates the experimental $T_{comp}$ of bulk TbIG [58].

The systematically lower $T_{comp}$ of TbIG thin films, despite various substrates, compared to the bulk value can be attributed to oxygen vacancies [11,32], which are generally present in garnet films [59,60]. We therefore further take into account the oxygen vacancy in our mean-field calculations. Here, we make a simple assumption that the superexchange interaction between the magnetic ions via an oxygen ion vanishes if the oxygen ion is absent. In bulk TbIG, each oxygen ion is shared by one $Fe^{O}$, one $Fe^{T}$, and two Tb ions. Introducing one such vacancy breaks the exchange between two pairs with $J_{ac}$, one pair with $J_{dc}$ and one pair with $J_{ad}$. We thus simulate oxygen vacancy effects by ignoring the field contributed by these exchange interactions. Recalculating $T_{comp}$ for bulk TbIG as a function of the oxygen vacancy amount, as shown in Fig. 5, we find a linear decay in $T_{comp}$. A few percent of oxygen vacancies leads to a decrease in $T_{comp}$ by tens of Kelvin. One can expect a certain amount of oxygen vacancies in TbIG/Pt films, as it was previously observed that $Fe^0$, instead of $Fe^{3+}$ or $Fe^{2+}$, accumulates at the interface between yttrium iron garnet and Pt films [61]. In addition, Fe depletion from garnet and diffusion into Pt [61] also lowers the field exerted on Tb ions and in turn reduces $T_{comp}$. The offsets of the vertical axis in Fig. 4(b) are accordingly justified.

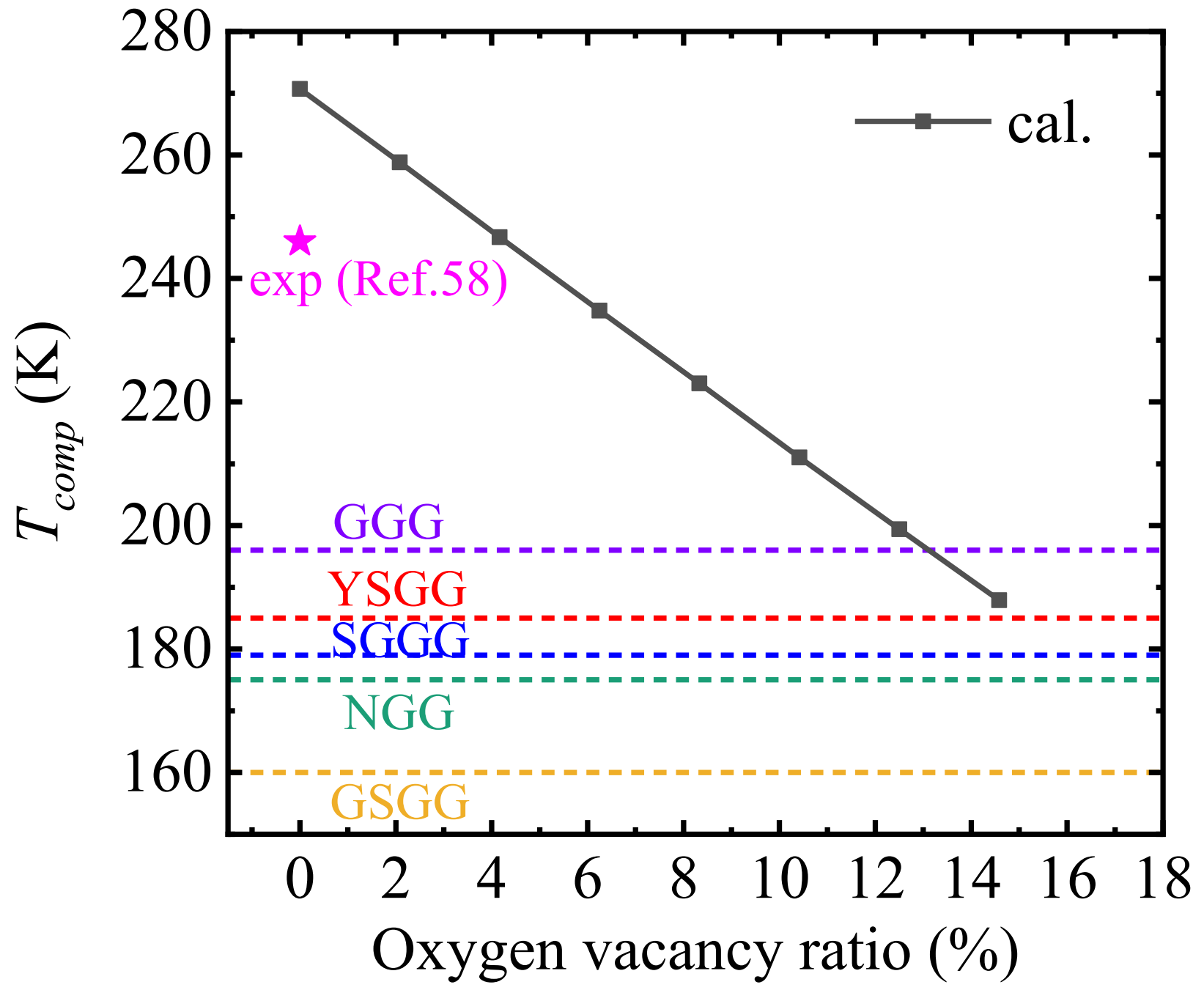

FIG. 5. Calculated compensation temperature of bulk TbIG as a function of the oxygen vacancy ratio. The dotted lines represent experimental values of TbIG thin films on various substrates. The pentagram denotes the measured value for bulk TbIG from [58].

**Conclusions:** In summary, we manipulate the lattice constant of ferrimagnet garnet TbIG thin films by choosing appropriate garnet substrates, which results in consecutive tuning of the in-plane strain. The XRD, RSM and STEM measurements confirm the exquisite epitaxial growth of TbIG films and imply a fully strained pseudomorphic structure with in-plane lattice constants equal to those of the substrates. By examining the sign change temperature of $R_{AH}$ in the TbIG/Pt heterostructure, we accurately determine the compensation temperatures of TbIG on these substrates with a precision down to 1 K. The compensation temperature monotonically decreases with increasing lattice constant. Density functional calculations show that the exchange interaction parameters between different magnetic sublattices exhibit a sensitive dependence on the lattice constant or bond length. This approach not only provides an accurate means of determining the compensation temperature of ReIG films but also enables modulation of the compensation temperature by means of strain. Our findings in this work show the significant potential of ferrimagnetic-insulator-based spintronics and practical applications.

Acknowledgements:
Y. L. and Z. S. would like to thank Lili Lang for her assistance on the STEM measurements. Work at Tongji University was supported by National Science Foundation of China Grants No. 12074285 and No. 11774259. Work at Beijing Normal University was supported by National Science Foundation of China Grant No.12174028.

[*]Yufei Li and Xihui Yang contributed equally to this work.
[†]Corresponding author.
yiliu@bnu.edu.cn

‡Corresponding author.
shizhong@tongji.edu.cn